\documentclass[12pt]{article}
\usepackage{graphicx}
\newcommand\pubnumber{SNSN-323-63}
\newcommand\pubdate{\today}

\def\Title#1{\begin{center} {\Large #1 } \end{center}}
\def\Author#1{\begin{center}{ \sc #1} \end{center}}
\def\Address#1{\begin{center}{ \it #1} \end{center}}

\newcommand\pubblock{\rightline{\begin{tabular}{l} \pubnumber\\
         \pubdate  \end{tabular}}}
\newenvironment{Abstract}{\begin{quotation}  }{\end{quotation}}
\newenvironment{Presented}{\begin{quotation} \begin{center} 
             PRESENTED AT\end{center}\bigskip 
      \begin{center}\begin{large}}{\end{large}\end{center} \end{quotation}}

\def\beq{\begin{equation}}
\def\eeq#1{\label{#1}\end{equation}}
\def\eeqn{\end{equation}}

\def\beqa{\begin{eqnarray}}
\def\eeqa#1{\label{#1}\end{eqnarray}}
\def\eeqan{\end{eqnarray}}

\let\bar=\overbar

\def\Dslash{\not{\hbox{\kern-4pt $D$}}}
\def\dslash{\not{\hbox{\kern-2pt $\del$}}}

\def\msb{{\bar{\ssstyle M \kern -1pt S}}}

\begin{document}
\begin{titlepage}
\pubblock

\vfill
\Title{X-ray Performance of Back-Side Illuminated Type of Kyoto's X-ray Astronomical SOI Pixel Sensor, XRPIX}
\vfill
\Author{Makoto Itou$^{1}$, Takeshi Go Tsuru$^{1}$, Takaaki Tanaka$^{1}$, Ayaki Takeda$^{1}$, 
Hideaki Matsumura$^{1}$, Shunichi Ohmura$^{1}$, 
Shinya Nakashima$^{2}$,\\Yasuo Arai$^{3}$, Koji Mori$^{4}$, Ryota Takenaka$^{4}$, Yusuke Nishioka$^{4}$, 
Takayoshi Kohmura$^{5}$, Koki Tamasawa$^{5}$, Craig Tindall$^{6}$}
\Address{$^{1}$Department of Physics, Graduate School of Science, Kyoto University,\\
	Kitashirakawa Oiwake-cho, Sakyo-ku, Kyoto 606-8502, Japan\\
$^{2}$Institute of Space and Astronautical Science (ISAS)/JAXA, 3-1-1 Yoshinodai, \\
	Chuo-ku, Sagamihara, Kanagawa 252-5210, Japan\\
$^{3}$Institute of Particle and Nuclear Studies, High Energy Accelerator Research Org., \\
	KEK, 1-1 Oho, Tsukuba 305-0801, Japan\\
$^{4}$Department of Applied Physics, Faculty of Engineering, University of Miyazaki,\\
	1-1 Gakuen Kibana-dai Nishi, Miyazaki 889-2192, Japan\\
$^{5}$Department of Physics, Faculty of Science and Technology, \\
	Tokyo University of Science, 2641 Yamazaki, Noda, Chiba 278-8510, Japan\\
$^{6}$Lawrence Berkeley National Laboratory, Berkeley, CA94720, USA}
\vfill
\begin{Abstract}
$\\$

We have been developing X-ray SOI pixel Sensors, called ``XRPIX'', for future X-ray astronomy satellites that enable us to observe in the wide energy band of 0.5$-$40 keV. 
Since XRPIXs have the circuitry layer with a thickness of about 8 $\mu$m in the front side of the sensor, it is impossible to detect low energy X-rays with a front-illuminated type. 
So, we have been developing back-illuminated type of XRPIX with a less 1 $\mu$m dead layer in the back-side, which enables the sensitivity to reach 0.5 keV. 
We produced two types of back-side illuminated (BI) XRPIXs, one of which is produced in ``Pizza process'' which LBNL developed and the other is processed in the ion implantation and laser annealing. 
We irradiated both of the BI-XRPIXs with soft X-ray and investigate soft X-ray performance of them. 
We report results from  soft X-ray evaluation test of the device.
%As a result, we find thickness of this dead layer is 2.0$^{+1.1}_{-0.6}$ $\mu$m. 
\end{Abstract}
\vfill
\begin{Presented}
International Workshop on SOI Pixel Detector\\(SOIPIX2015)\\
Tohoku University, Sendai, Japan, 3-6, June, 2015
\end{Presented}
\vfill
\end{titlepage}
\def\thefootnote{\fnsymbol{footnote}}
\setcounter{footnote}{0}
\section{Introduction}
~~~Charge-coupled devices (CCDs) are standard imaging spectrometers widely used in modern X-ray astronomy. 
They have a good performance of spectroscopy and imaging because of their small pixel sizes (20 $\mu$m $\sim$ 30 $\mu$m) 
and the Fano-factor limitation on the energy-resolution ($\sim$ 130  eV in FWHM at 6 keV) \cite{G.P.Garmire}-\cite{K.Koyama}. 
However, CCDs suffer from the problem such as a high non-X-ray (NXB) background above 10 keV induced by the interactions with cosmic-rays in orbit. 
One of ways of solving this problem is adopting an anti-coincident technique with surrounding scintillators as active shields\cite{T.Takahashi}\cite{T.Anada}.
On the other hand, CCDs do not have a good time-resolution which is enough to adopt this technique.
Thus, we have been developing X-ray SOI pixel sensors, called ``XRPIX'',  which overcome the limitations of CCDs, in order to realize wide-band X-ray imaging spectroscopies for the future X-ray astronomy\cite{T.G.Tsuru}.
XRPIXs enable us to observe with a high time-resolution of $\sim$10 $\mu$sec, which allows us to employ anti-coincidence shields and reject NXB.

\begin{figure}[h]
\centering
\includegraphics[height=2in]{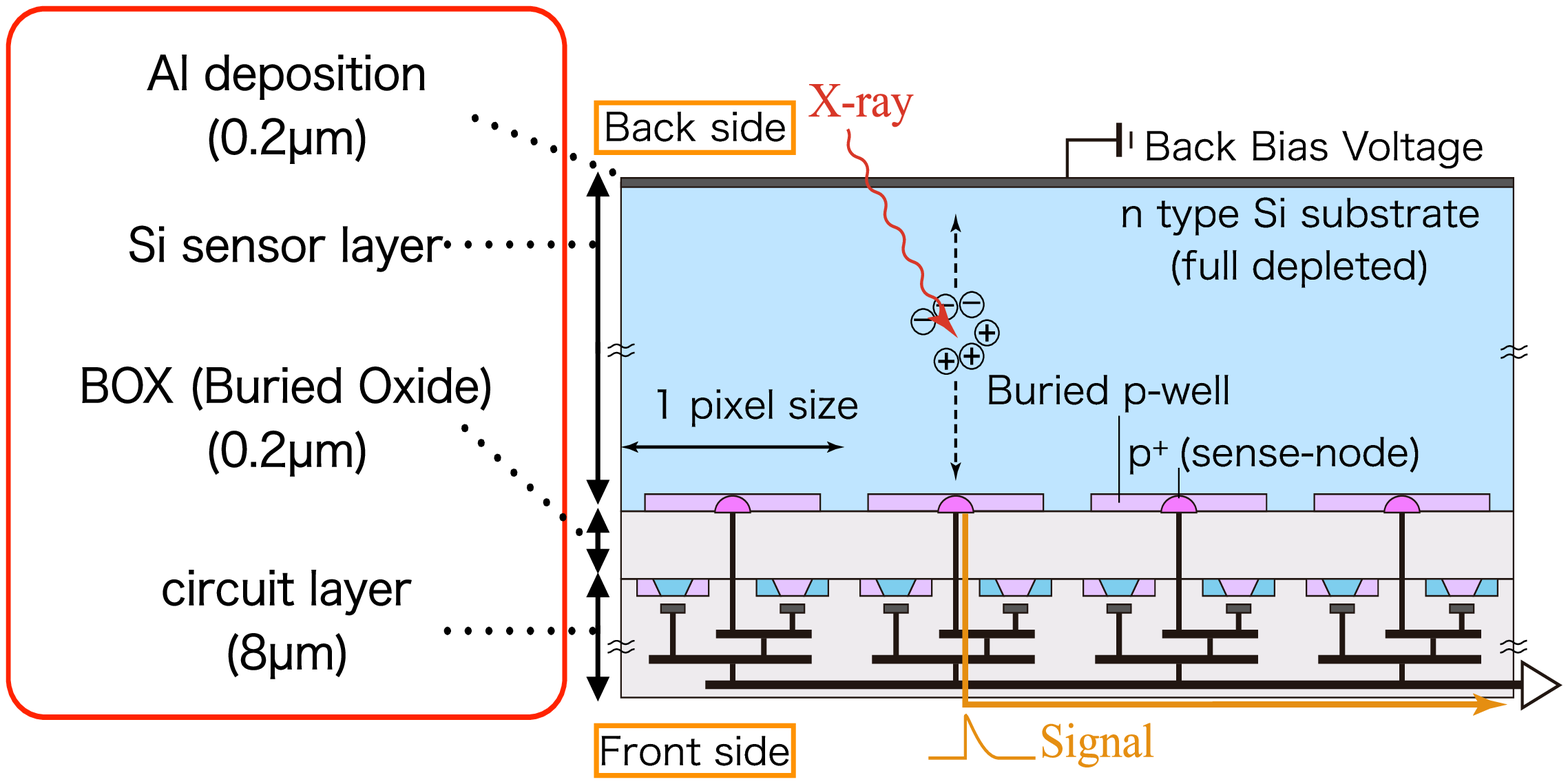}
\caption{The cross-sectional view of XRPIX.}
\label{fig:XRPIX2b}
\end{figure}

XRPIXs are the monolithic active pixel sensors, which are processed with the silicon-on-insulator (SOI) CMOS technology\cite{Y.Arai}. 
As Figure \ref{fig:XRPIX2b} shows,  an XRPIX consists of three layers: a circuit with low-resistivity silicon ($\sim$ 8 $\mu$m), a sensor with high-resistivity silicon, and SiO$_{2}$ insulator between two. 

In order to detect X-rays with the energies above $\sim$ 10 keV, we need to develop the XRPIX with the Si thick sensor-layer.
We need to use the back-side of the XRPIX (back-illumination, BI) as an entrance face for incident X-rays because a circuitry layer on the front-side of it prevents us from detecting low-energy X-rays blow 1 keV.
A dead layer on the X-ray incident surface should ideally be as thin as possible to achieve a high
sensitivity to low-energy X-rays.
Thus, the fully-depleted BI-type of XRPIX with a thin dead layer is required to realize the wide bandpass performance.
In this paper, we report the X-ray performance of the first BI-types of XRPIXs, ``XRPIX-FZ-LA'' and ``XRPIX-CZ-PZ''.

%10keV以上の高いエネルギーX線を検出するには，厚い空乏層が必要である．
%It is impossible to detect low energy X-rays with front-illuminated type because XRPIX has the circuitry layer with thickness of about 10 $\mu$m in the front side of sensor. 
%そこで，XRPIXでは厚い完全空乏層(eg. $>200$um)と1um以下の薄い不感層の両立を目指す(Fig.1)．
%So, we have been developing back-illuminated type of XRPIXs with a less 1 $\mu$m dead layer in the back-side, which enables the sensitivity to reach 0.5 keV. 

%空乏層が厚くなるとcharge diffusionにより，charge sharingしたイベントが多くなる．
%X線イベントの電荷量を再構成するために，複数のピクセルデータを加算する必要がでる．
%その結果，実質的に読み出しノイズが大きくなりエネルギー分解能を劣化させる．
%さらに，実効的に低エネルギーX線の閾値が高くなる．
%一方，軟X線感度を決める裏面構造として，暗電流が少なく，不感層が薄いことが必要である．
%さらには高い分光性能を得るために，裏面付近のCCEが高いことが必要である．
%従って，性能の高いワイドバンド性能を得るには，
%高い比抵抗を持つシリコン素材の開発 which determines デバイスの厚み，薄い不感層を実現する裏面の構造，
%低い読み出しノイズを開発し，ピクセルサイズ，高いバックバイアスを印加可能なガードリングなど多くの要素で，
%努力をする必要がある．

%この論文では，その努力の最初として，厚い裏面照射型の最初のプロトタイプとして製造した素子"XRPIX2b-FZ-LA"と，
%裏面感度の改善を試みた素子"XRPIX2b-Pizza"のX線性能を報告する．
% We produced XRPIX2b-CZ-Pizza and XRPIX2b-FZ-Lapis as back-illuminated XRPIX. 
% The one is produced in "Pizza process"\cite{M.Battaglia} which LBNL has developed, 
% and the other is processed in the ion implantation and laser annealing by Lapis. 
% In this study, we irradiate this sensors with soft X-ray and investigate the performance of them.

\section{Experimental Setup}
\begin{figure}[h]
\centering
\includegraphics[height=6cm]{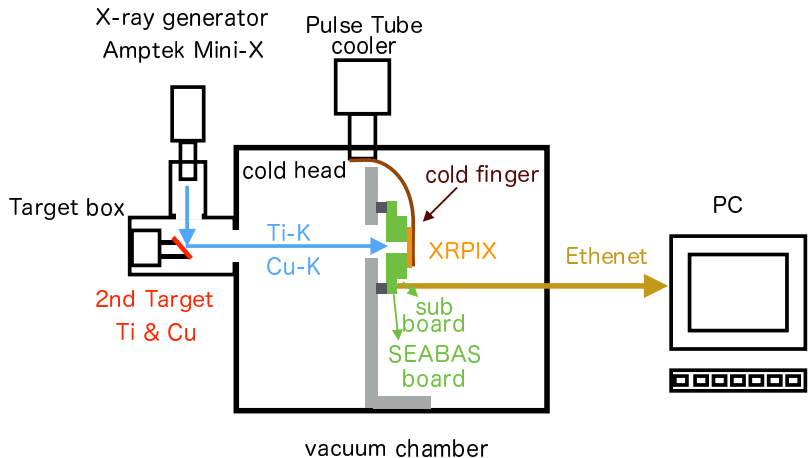}
\caption{The picture of the experimental setup.}
\label{fig:setup}
\end{figure}

~~~
We irradiated XRPIX with Ti-K and Cu-K. 
Ti-K (4.51 keV) and Cu-K (8.04 keV) lines have attenuation lengths of 13.1 $\mu$m and 67.3 $\mu$m, respectively. 
Figure \ref{fig:setup} shows the schematic of the experimental setup in the vacuum chamber. 
%We irradiated a silicon drift detector (Amptek XR-100) or XRPIX sensor with the X-ray generator (Amptek mini-X). 
% Ti-K and Cu-K X-ray fluorescent lines are illuminated at the same time on the SDD or XRPIX's back-side window. 
XRPIX is connected to the cold head and cooled by Iwatani pulse tube cooler (Iwatani CryoMini P003) to cool the XRPIX to $-50^{\circ}$C.
We use a readout system consisting of a sub board with which an XRPIX is equipped, and  a SEABAS (Soi EvAluation BoArd with Sitcp) board. 
Details of the readout system are reported in \cite{T.Uchida}.

\begin{figure}[h]
\centering
\includegraphics[height=6cm]{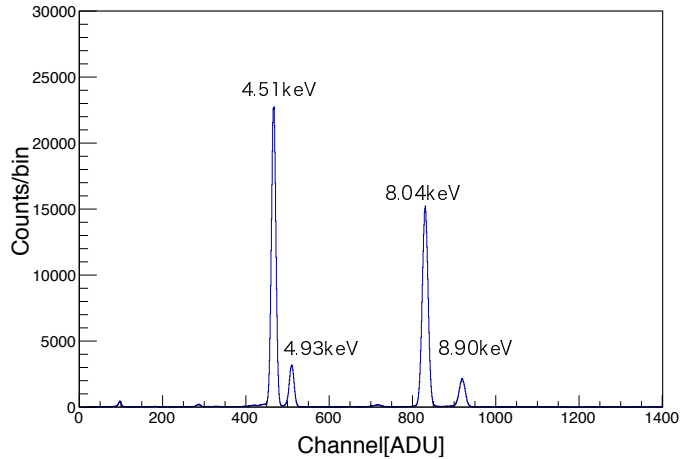}
\caption{The spectrum obtained by SDD.}
\label{fig:sdd}
\end{figure}

Before the performance evaluation of XRPIXs, we observed the spectra and absolute flux of the Ti-K and Cu-K lines with a silicon drift detector (SDD: Amptex XR-100).
%実験を行う前に，Ｘ線スペクトルと絶対強度のキャリブレーションをamptex sddを用いて行った．
Figure \ref{fig:sdd} shows the spectrum of illuminated X-ray obtained with the SDD. 
Ti-K lines at 4.51 keV and 4.93 keV and Cu-K lines at 8.04 keV and 8.90 keV are seen. 
%単色性は今回の実験の精度に十分に良い．
We obtained absolute fluxes of the lines at the position of the XRPIX, referring to the quantum efficiency and the collimator size of the SDD given in the Amptek website\cite{Amptek}, and correcting the difference in the distances from the secondary target to the XRPIX and SDD.
The errors in the absolute fluxes are dominated by the systematic error of the collimator area of the SDD ($\pm$10$\%$).
%絶対強度のエラーはSDDの窓のコリメーターの面積の工作精度によるエラー（±10\%）が最も大きい．
%以下に示す実験結果にはそのエラーを取り込んでいる．

%\begin{figure}[ht]
%\begin{tabular}{p{0.45\textwidth}p{0.1\textwidth}p{0.45\textwidth}}
%\centering\includegraphics*[ height=1.6in, clip]{/soipix2015/imagesetup_v1.eps}
%\caption{The picture of the experimental setup.}
%\label{fig:setup}
%&&
%\centering\includegraphics*[ height=1.6in, clip]{/soipix2015/imageSDD.eps}
%\caption{The spectrum obtained by SDD.}
%\label{fig:sdd}
%\end{tabular}
%\end{figure}

\section{XRPIX-FZ-LA}

\subsection{Device Discription}
~~~
We produced two BI-XRPIXs, named ``XRPIX-FZ-LA'' and ``XRPIX-FZ-PZ''. 
So, we introduce specifications of them, which are summarized in table \ref{tab:DeviceSpec}.
We used the 0.2-$\mu$m CMOS fully-depleted SOI CMOS Pixel process provided by Lapis Semiconductor Co. Ltd..
Two BI-XRPIXs are different in terms of the resistivity in the Si sensor-layers.
One is the Floating Zone (FZ) device with nominal resistivity is $\sim$ 4 k$\Omega \cdot$cm, 
and the other one is the Czochralski (CZ) device with nominal resistivity is $\sim$ 1.2 k$\Omega \cdot$cm.
The XRPIX-FZ-LA has the FZ-type Si sensor-layer with the thickness of 500 $\mu$m.
On the back-side of a sensor layer, chemical etching, ion implantation, laser annealing, and vapor-deposition of aluminum (200 nm) for optical blocking are processed.

\begin{table}[thb]
	\begin{center}
		\caption{Specifications of XRPIX-FZ-LA and -CZ-PZ}
  		\begin{tabular}{ccc}\hline
                                     & XRPIX2b-FZ-LA &  XRPIX2b-CZ-PZ\\ \hline
		   Pixel size ($\mu$m $\times$ $\mu$m) & \multicolumn{2}{c}{30 $\times$ 30}\\
		   Format (pixels)  &  \multicolumn{2}{c}{152 $\times$ 152}\\
		   Wafer type & Floating Zone & Czochralski \\
		   Nominal Resistivity (k$\Omega \cdot$cm) & $\sim$ 4 &  $\sim$ 1.2 \\
		   Sensor layer ($\mu$m) & 500 & 62 \\
		   Back-side process & ion implantation, & Pizza process \\
		                     & laser annealing & \\
		   Al deposition & 0.2 $\mu$m & none \\ \hline
		\end{tabular}
		\label{tab:DeviceSpec}
	\end{center}
\end{table}

\subsection{Experiments and Results}
~~~
%There are two types of XRPIX2b (CZ-type and FZ-type) by difference of the resistivity Si sensor layer. CZ-type (FZ-type) usually has the 260$\mu$m (500$\mu$m) sensor layer and resistivity of 1.5k$\Omega \cdot$cm (5k$\Omega \cdot$cm). 
%XRPIX2b dtails are repoted in \cite{A.Takeda}.  
%Table \ref{tab:device} shows two type XRPIXs of details.
%XRPIX2b-FZ-LA is processed in the ion implantation and laser annealing by Lapis.
%その後，遮光アルミニウム0.2umをデポジットとしてある．
%下記に示すＸ線実験では，その厚みを含んでいることに注意する．
With the XRPIX-FZ-LA, 
we obtained spectra and the flux of Ti-K and Cu-K with back-bias voltages (V$_{bb}$) from 140 V to 300 V.
% and 140V $\sim$ 300V by XRPIX2b-CZ-Pizza and XRPIX2b-FZ-Lapis, respectively.  
The chip was cooled to -40 $^{\circ}$C $\sim$ -30 $^{\circ}$C at the degree of vacuum lower than 10$^{-5}$ Torr. 
We performed a frame-by-frame readout, where all the pixels are sequentially read out after a 1 ms exposure. 
The total exposure time is 300 s, respectively. 
The details of the readout sequence is reported in \cite{Ryu}.
In order to classify charge sharing events, we used the methods described in \cite{Ryu} and \cite{Nakashima}. 
We picked up a pixel whose pulse height exceeds a predefined threshold called event threshold. 
Then, we checked pulse height of adjacent 8 pixels whether they exceed a split threshold and classified each X-ray event into four event types of ``single pixel'', ``double pixel'', ``triple pixel'', and ``other''.
%We defined the event and split threshold as 10s and 5s of readout noise, respectively. 
%When we select events from the raw data, we use two pulse height threshold. 
%One is the event threshold at 8 sigma and the other is the split threshold at the 4 sigma significance with respect  to the noise level. 
In this analysis, the readout noise is 59 e$^{-}$, the event threshold is 2.4 keV, and the split threshold is 0.5 keV. 
% \subsection{Spectral Performance}

\begin{figure}[h]
\centering
\includegraphics[height=5cm]{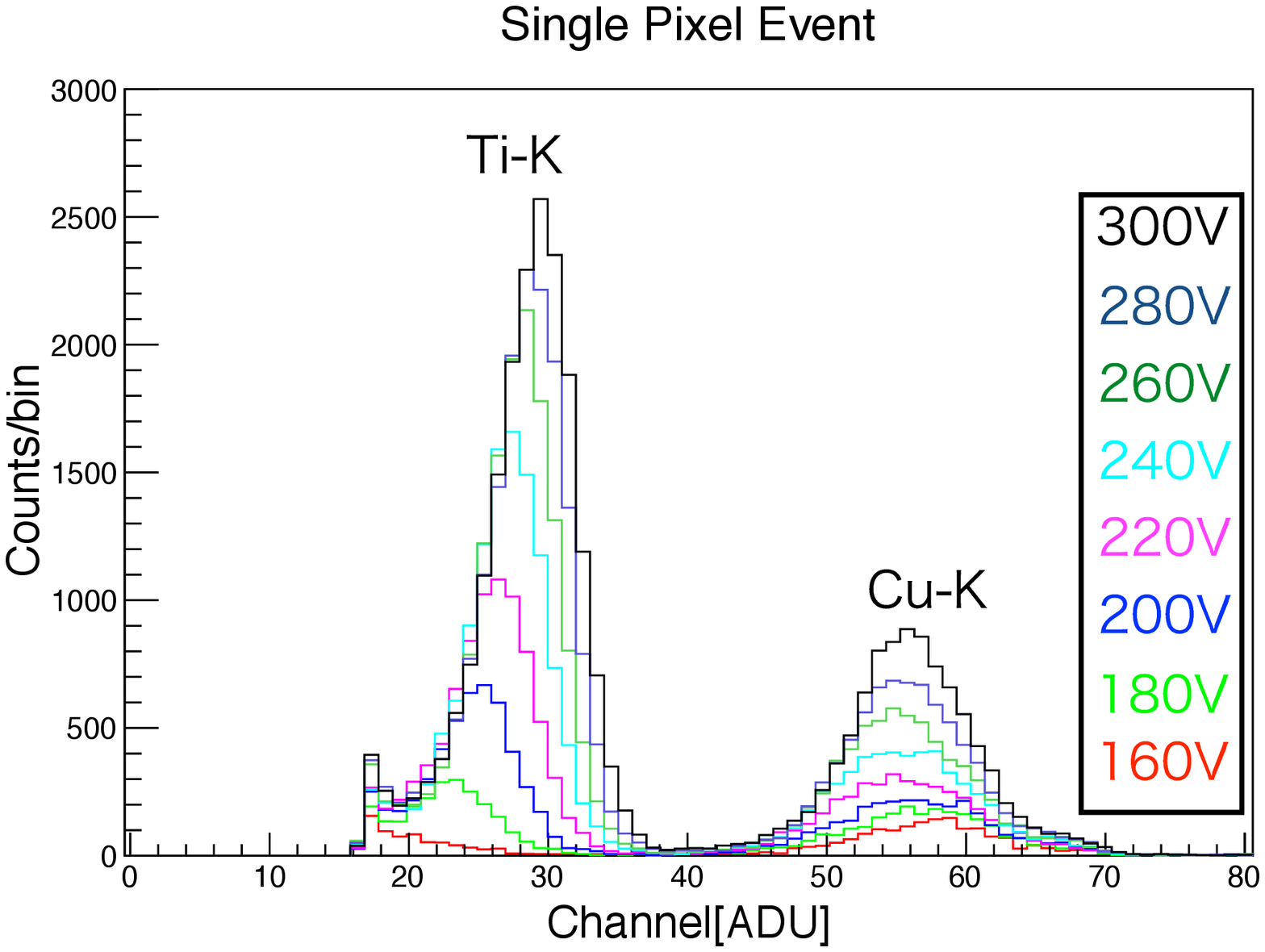}
\caption{The spectra of single pixel events obtained for various back bias voltages with XRPIX-FZ-LA. }
\label{fig:spectrum_lapis}
\end{figure}

\begin{figure}[ht]
\begin{tabular}{p{0.45\textwidth}p{0.1\textwidth}p{0.45\textwidth}}
\centering\includegraphics*[ height=1.6in, clip]{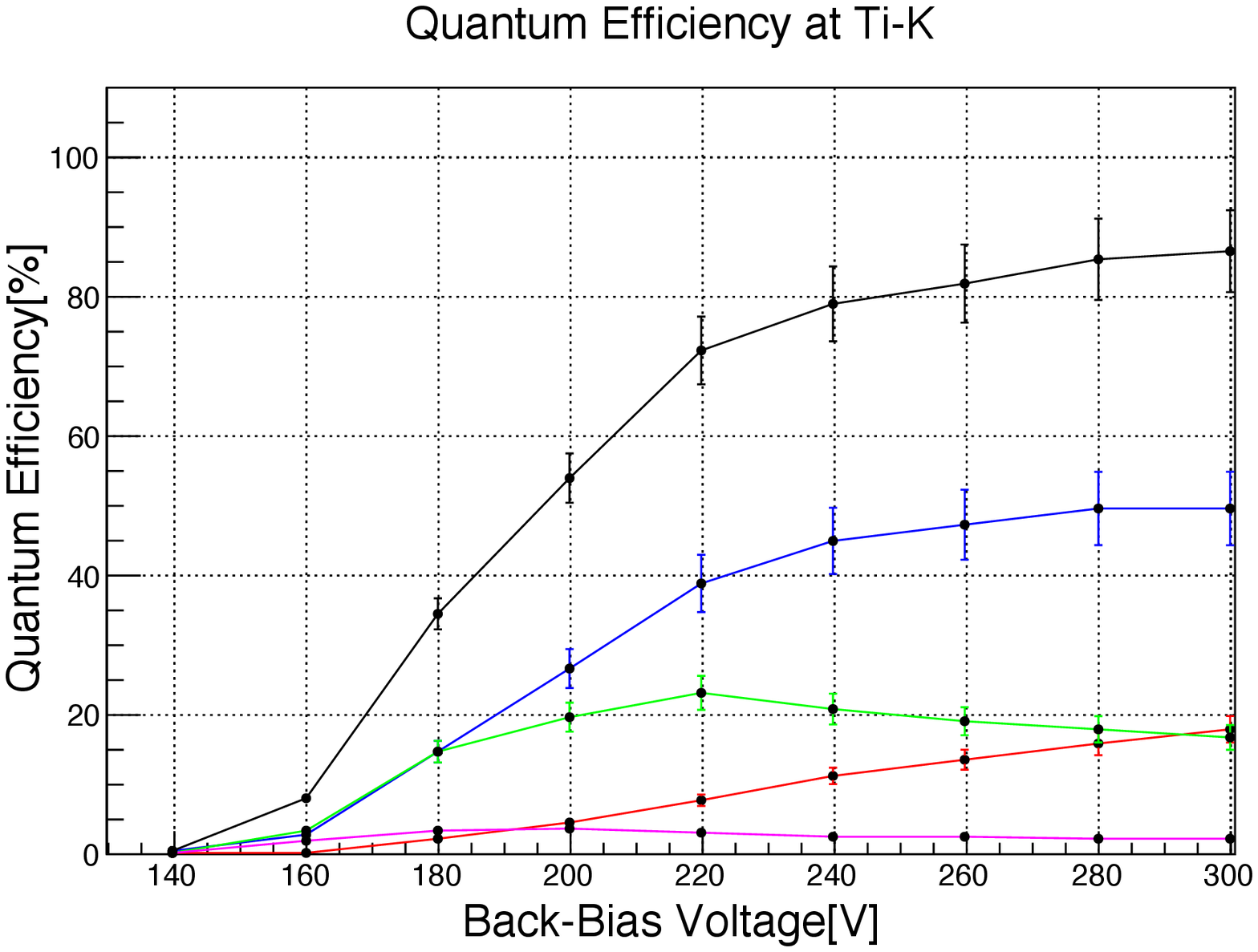}
\caption{Quantum efficiency of each grade with XRPIX-FZ-LA at the energy of Ti-K X-rays as a function of V$_{bb}$}
\label{fig:qe-lapis-Ti}
&&
\centering\includegraphics*[ height=1.6in, clip]{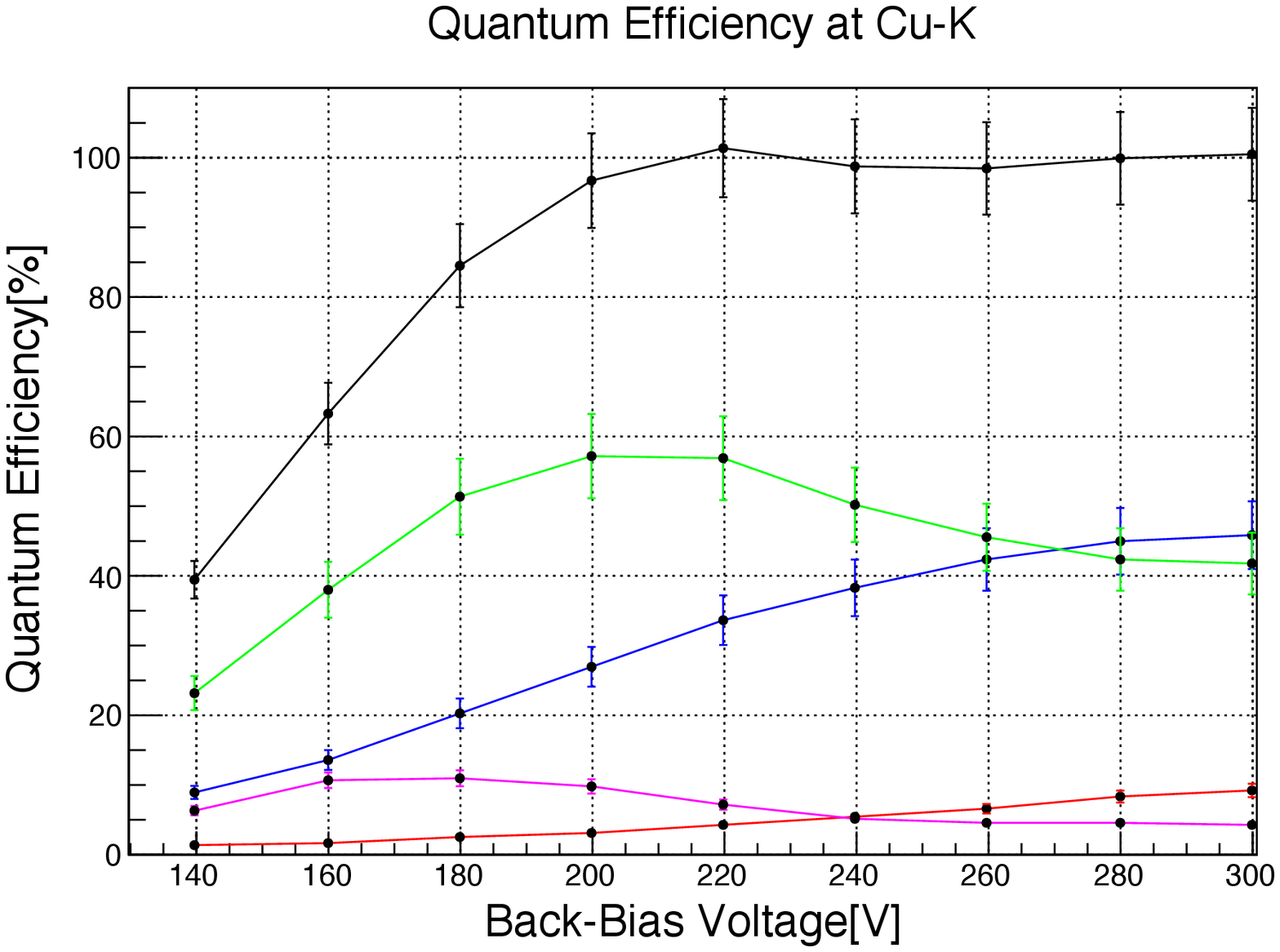}
\caption{Same as Figure~\ref{fig:qe-lapis-Ti} but at the energy of Cu-K X-rays.}
\label{fig:qe-lapis-Cu}
\end{tabular}
\end{figure}

Figure \ref{fig:spectrum_lapis} shows spectra obtained by XRPIX-FZ-LA with back bias voltages from 160V to 300V. 
Figure \ref{fig:qe-lapis-Ti} and  \ref{fig:qe-lapis-Cu} shows the quantum efficiency of each grades separately as a function of the applied back-bias voltage.
These figures indicate that count rates of Ti-K and Cu-K increase.
When we changed back V$_{bb}$ from 160V to 300V, the ratio of Ti-K counts becomes 11.11. 
Similarly, one of Cu-K counts becomes 1.58.
The growth rate of counts is clearly different between Ti-K and Cu-K.
From this result, we understand the depletion layer in the Si sensor-layer spreads from the front-side of face to the back-side of one.
The QE with the sum of four grades become constant at V$_{bb}$ $\geq$ 220V.
The QE of Cu-K line is consistent with the expected value for absorption in the Si sensor-layer with the thickness of 500 $\mu$m. 
This results suggest the Si sensor-layer gets fully depleted at $\sim$ 220 V.

\begin{figure}[h]
\centering
\includegraphics[height=5cm]{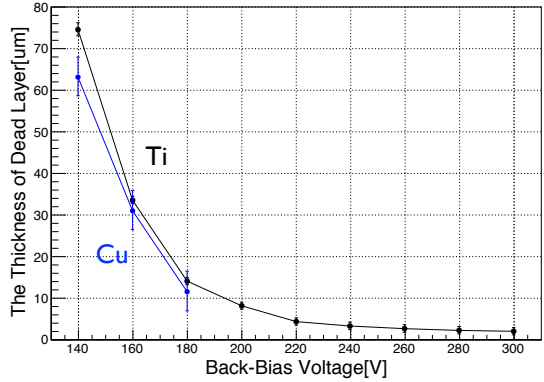}
\caption{The Thickness of Dead Layer@XRPIX-FZ-LA}
\label{fig:deadlayer_lapis}
\end{figure}

Figure \ref{fig:deadlayer_lapis} shows the thickness of the dead layer which we estimated from the QEs of Ti-K and Cu-K as a function of V$_{bb}$.
Ti-K line's attenuation length is shorter than Cu-K's one.
So, the dead layer at V$_{bb}$ $\geq$ 220 V is also obtained from the QEs of Ti-K in oder to estimate it accurately.
Before the Si sensor-layer gets fully depleted, the dead layer is thicker.
So, we estimated the thickness of the dead layer at V$_{bb}$ $<$ 220 V from the QE of Cu-K.
From these results, the thickness of the dead layer is 1.9 $\pm$ 0.9 $\mu$m which includes the Al deposition layer of 0.2 $\mu$m thick.

According to the above, XRPIX-FZ-LA is a full depleted device, but its dead layer is not satisfied with the requirement of 1 $\mu$m.
%The error of these experiments are large, and the thickness of the dead layer is not accurately estimated.
In oder to achieve the higher sensitivity to low-energy X-rays, we  need to improve the device.

We discuss whether the thickness of the Si sensor-layer is adequate to detect low-energy X-rays. 
The sensor-layer with thickness of 500 $\mu$m is thick enough to observe in the high energy X-ray band.
On the other hand, the single pixel events don't have a small proportion of the sum of X-ray events due to the thick sensor-layer.
In other words, the charge clouds expand along the electric flux line, which are shared by multi pixels.
This phenomenon is seen in Figure \ref{fig:qe-lapis-Ti} and \ref{fig:qe-lapis-Cu}.
Since charge collection efficiency is better as higher V$_{bb}$ is applied, the charge of Ti-K is smaller, and the triple and other pixel events decrease.
According to Figure \ref{fig:qe-lapis-Ti} and \ref{fig:qe-lapis-Cu}, the single pixel event is about 10 $\sim$ 20 $\%$ of the charge.
The large number of multi pixel events make the readout noise increase effectively.
In order to achieve the good performance of spectroscopy enough to observe low-energy X-rays and estimate the thickness of dead layer, the sensor-layer is thinner than 500 $\mu$m.
So, we developed the device with the thinner Si sensor-layer which is produced by new back-side process.

\section{XRPIX-CZ-PZ}
~~~We fabricated the devices, ``XRPIX2b-CZ-PZ'' ,which has a thin Si sensor-layer of 62 $\mu$m thick  and the back-side treated with the “Pizza process”, developed by LBNL\cite{M.Battaglia}.
The specifications of the device are given in Table \ref{tab:DeviceSpec}.
Pizza process" is the way of processing the back-side widow and goes through the following procedure.
First, the Si wafer is back-thinned to 70 $\mu$m by using grinding technique. 
Second, a thin phosphor layer is made in the wafer by ion-implantation at -160 $^{\circ}$C. 
Finally, it is annealed at 500 $^{\circ}$C for 10 min like baking a pizza. 
The details of the process is give in \cite{M.Battaglia}.

\begin{figure}[h]
\centering
\includegraphics[height=5cm]{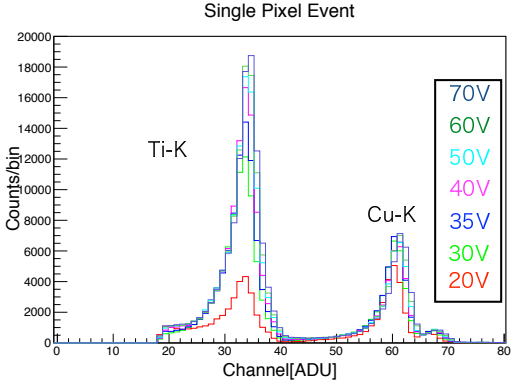}
\caption{The spectrum obtained by XRPIX2b-CZ-Pizza}
\label{fig:spectrum_pizza}
\end{figure}

\begin{figure}[ht]
\begin{tabular}{p{0.45\textwidth}p{0.1\textwidth}p{0.45\textwidth}}
\centering\includegraphics*[ height=1.6in, clip]{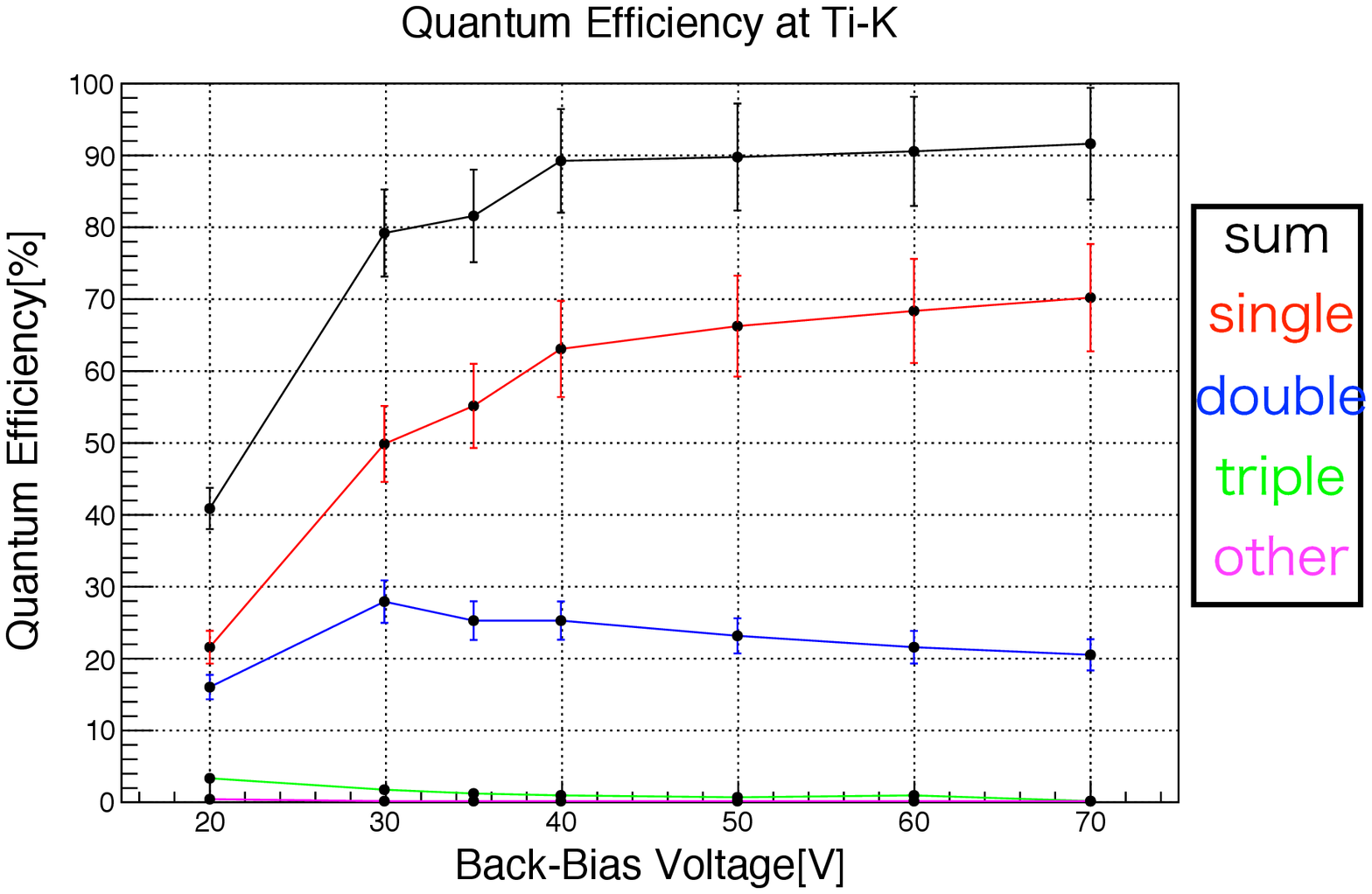}
\caption{Quantum efficiency of each grade with XRPIX-CZ-PZ at the energy of Ti-K X-rays as a function of V$_{bb}$}
\label{fig:qe-pizza-Ti}
&&
\centering\includegraphics*[ height=1.6in, clip]{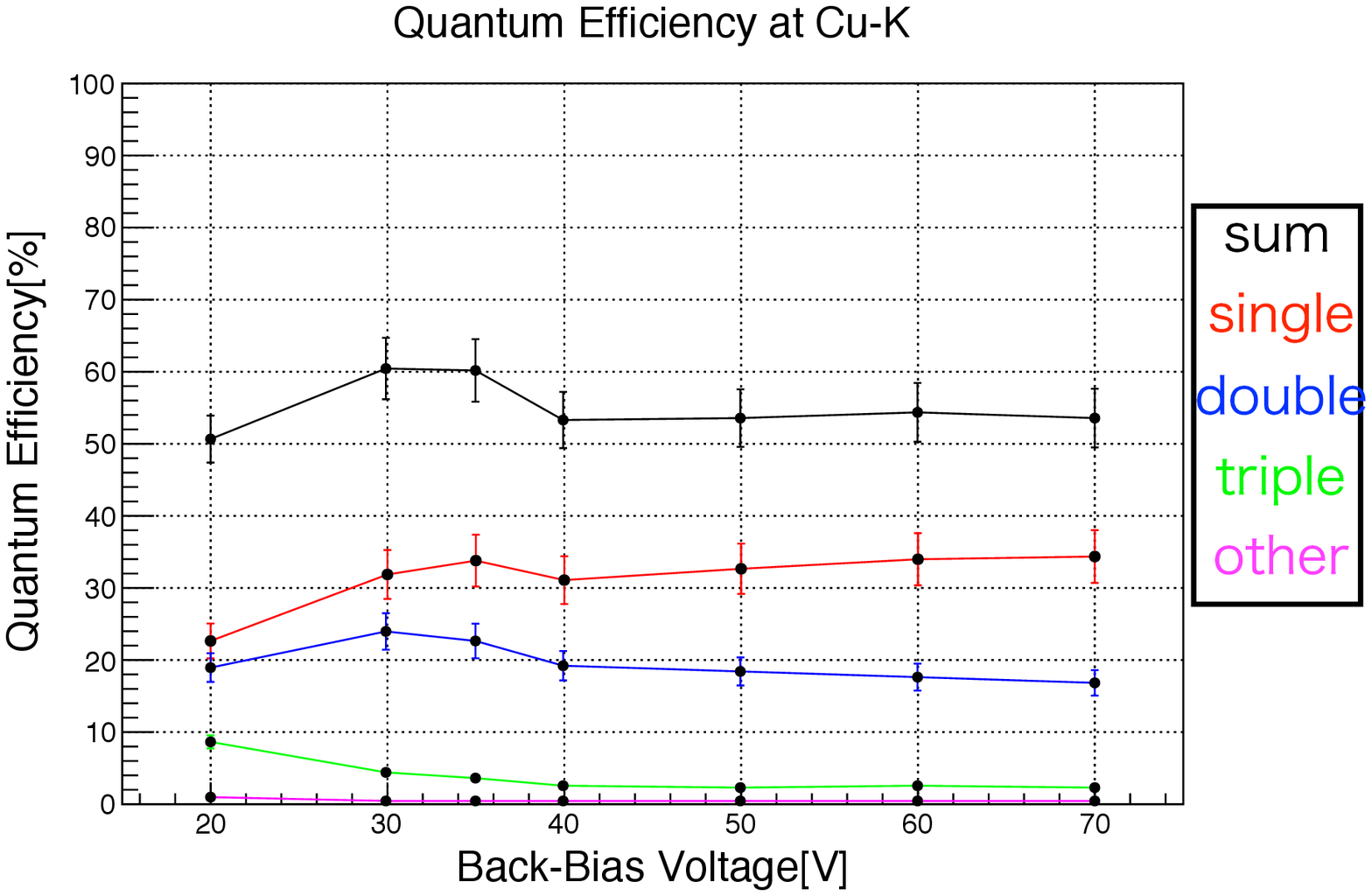}
\caption{Same as Figure~\ref{fig:qe-pizza-Cu} but at the energy of Cu-K X-rays.}
\label{fig:qe-pizza-Cu}
\end{tabular}
\end{figure}

Figure \ref{fig:spectrum_pizza} shows that the spectra obtained by XRPIX-CZ-PZ with back bias voltages of 20V $\sim$ 70V.  
In this analysis, the readout noise is 44 e$^{-}$, and we use the same event and split thereshold as XRPIX-FZ-LA. 
We can see Cu-K$\alpha$ and Cu-K$\beta$ separated from Cu-K line. 
The spectra with different back bias voltages are completely different. 
As higher back bias voltage is applied, the event of Ti-K increases exactly. 
Figure \ref{fig:qe-pizza-Ti} and \ref{fig:qe-pizza-Cu} show the QE of each grades as a function of the V$_{bb}$.
According to these figures, the ratio of Ti-K counts becomes 2.24 when we change back bias voltage from 20V to 70V.
Similarly, one of Cu-K counts becomes 1.06. 
From this result, we can understand the dead layer is thinner as the depletion region spreads.
The QE with the sum of four grades become constant at V$_{bb}$ $\geq$ 40V.
The QE of Cu-K line is consistent with the expected value for absorption in the Si sensor-layer with the thickness of 62 $\mu$m. 
This results suggest the Si sensor-layer gets fully depleted at $\sim$ 40 V.
The QE of Ti-K increases as the back bias voltage that is higher and becomes almost constant when $\geq$ 40V. 

\begin{figure}[htbp]
\centering
\includegraphics[height=5cm]{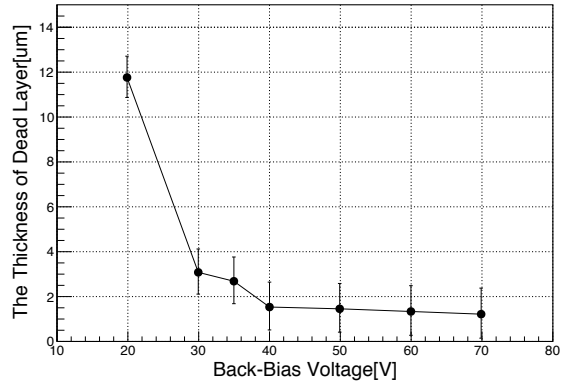}
\caption{The Thickness of Dead Layer@XRPIX-CZ-PZ}
\label{fig:deadlayer_pizza}
\end{figure}

Figure \ref{fig:deadlayer_pizza} shows the thickness of the dead layer which we estimated from the QEs of Ti-K and Cu-K as a function of V$_{bb}$.
From this results, the thickness of dead layer is 1.2 $\pm$ 1.1 $\mu$m. 
The thickness of dead layer nearly satisfies the requirement for future X-ray satellite. 
This large error suggests that the attenuation length of Ti-K is long enough to estimate the thickness of the dead layer.
So, we need to estimate the thickness of the dead layer with a fluorescent X-ray with the shorter attenuation length than the one of Ti-K.
We have developed the

% \newpage
\section{Summary and Future Plan}
~~~We have been developing back-illuminated types of XRPIXs in order to have good sensitivity down to 0.5 keV for future satellite. 
We evaluate the thickness of the performance for low-energy X-rays and dead layer of XRPIX-FZ-LA and XRPIX-CZ-PZ. 
The dead layer thickness of XRPIX-FZ-LA is 1.9 $\pm$ 0.9 $\mu$m, and the one of XRPIX-CZ-PZ is 1.2 $\pm$ 1.1 $\mu$m.
These result suggest the thickness of the dead layer nearly satisfies the requirement for a future X-ray satellite.
In the next experiments, we will evaluate the dead layer thickness and Charge-Collection-Efficiency (CCE) etc. by using lower energy X-rays such as Cl-K and Al-K line.

\end{document}